\documentclass[12pt,oneside]{article}
\usepackage{epsfig, cite}
\usepackage{color}
\usepackage{ifthen}
\usepackage{graphicx}
\usepackage{amsmath}

\def\p{\partial}

\def\half{{1\over 2}}
\def\({\left(}
\def\){\right)}
\def\[{\left[}
\def\]{\right]}

\def\e{\begin{equation}}
\def\q{\end{equation}}
\def\m{\begin{eqnarray}}
\def\n{\end{eqnarray}}

\begin{document}
\thispagestyle{empty} \setcounter{page}{0}
\renewcommand{\theequation}{\thesection.\arabic{equation}}

\begin{flushright}

\hfill{USTC-ICTS-08-18}\\

\end{flushright}
\vspace{1cm}

\begin{center}
{\huge Curvaton Dynamics and the Non-Linearity Parameters in
Curvaton Model}

\vspace{1.4cm}

Qing-Guo Huang$^1$ and Yi Wang$^{2,3}$

\vspace{.2cm}

{\em $^1$School of physics, Korea Institute for
Advanced Study,} \\
{\em 207-43, Cheongryangri-Dong,
Dongdaemun-Gu, } \\
{\em Seoul 130-722, Korea}\\
\vspace{.2cm} {\em $^2$Institute of Theoretical Physics, CAS,}\\{\em
Beijing 100080, P.R.China} \\\vspace{.2cm} {\em $^3$The
Interdisciplinary Center for Theoretical Study, USTC,}\\{\em Hefei,
Anhui
230026, P.R.China, } \\

\vspace{.2cm}

\centerline{{\tt huangqg@kias.re.kr}}

\centerline{{\tt wangyi@itp.ac.cn}}

\end{center}

\vspace{0.5cm}

\centerline{ABSTRACT}
\begin{quote}

We investigate the curvaton dynamics and the non-linearity
parameters in curvaton model with potential slightly deviating from
the quadratic form in detail. The non-linearity parameter $g_{NL}$
will show up due to the curvaton self-interaction. We also point out
that the leading order of non-quadratic term in the curvaton
potential can be negative, for example in the axion-type curvaton
model. If a large positive $g_{NL}$ is detected, the axion-type
curvaton model will be preferred.

\end{quote}
\baselineskip18pt

\noindent

\vspace{5mm}

\newpage

\setcounter{equation}{0}
\section{Introduction}

Inflation model \cite{Guth:1980zm} provides an elegant mechanism to
solve many puzzles in the hot big bang model due to the
quasi-exponential expansion of the universe. The quantum fluctuation
generated during inflation naturally seeds the small anisotropies in
the cosmic microwave background radiation and the formation of the
large-scale structure. In general many scalar fields are present in
the early universe. The dynamics of inflation is governed by
inflaton(s), but the primordial curvature perturbation can be
generated by inflaton(s) \cite{Guth:1982ec} or curvaton(s)
\cite{Linde:1996gt}, or both of them. However the feature of the
primordial power spectrum caused by inflaton is similar to curvaton.

The CMB non-Gaussianity \cite{Komatsu:2001rj} opens a new window to
probe the physics of the early universe. A well-discussed ansatz of
non-Gaussianity has the local shape. This kind of non-Gaussianity
can be characterized by some non-linearity parameters:
\begin{equation}
  \zeta({\bf x})=\zeta_g({\bf x})+\frac{3}{5}f_{NL}\zeta_g^2({\bf x})+\frac{9}{25}g_{NL}\zeta_g^3({\bf x})+\cdots~,
\end{equation}
where $\zeta$ is the curvature perturbation in the uniform density
slice, $\zeta_g$ denotes the Gaussian part of $\zeta$. The
non-linearity parameters $f_{NL}$ and $g_{NL}$ parameterizes the
non-Gaussianity from the irreducible 3-point and 4-point correlation
functions respectively.

In the case of single field inflation model $f_{NL}^{local}\sim
{\cal O}(n_s-1)$ \cite{Maldacena:2002vr}, which is constrained by
WMAP ($n_s=0.960_{-0.013}^{+0.014}$) \cite{Komatsu:2008hk} to be
much less than unity. In some general inflation models
\cite{Chen:2006nt} a large non-Gaussianity is possibly obtained, but
the shape of non-Gaussianity is different from the local shape. As
we know, curvaton model can easily generate a large local-type
non-Gaussianity \cite{Linde:1996gt,Lyth:2002my}. Different group
reported different result of data analysis to $f_{NL}^{local}$ based
on WMAP 3yr data: \e 27<f_{NL}^{local}<147 \q in \cite{Yadav:2007yy}
and \e -70<f_{NL}^{local}<91 \q in \cite{Hikage:2008gy} at $95\%$
CL. WMAP group reported the 5yr result \cite{Komatsu:2008hk}:
\begin{equation}\label{fnlexp}
  -9<f_{NL}^{local}<111 \quad (95\% \ \hbox{CL}).
\end{equation}
Up to now a Gaussian distribution of the primordial curvature
perturbation is still consistent with experiments. However, the
lower bound on $f_{NL}^{local}$ in Eq.\eqref{fnlexp} is greatly
reduced from WMAP3 to WMAP5. If a large positive $f_{NL}^{local}$ is
confirmed by more years integration of WMAP, or the Planck mission,
it can help us to distinguish curvaton model from inflation model.
Recently many issues related to curvaton were widely discussed in
\cite{Huang:2008ze,Huang:2008rj,Enqvist:2008gk}.

Curvaton is a light scalar field with weak self-interaction. Even
though its energy density is subdominant during inflation, it can
make the main contribution to the primordial curvature perturbation.
In curvaton model, $f_{NL}$ is inversely proportional to the
fraction of curvaton energy density in the energy budget at the
epoch of curvaton decay, and the value of $g_{NL}$ is mainly
contributed by the curvaton self-interaction. In the literatures,
the interaction term of curvaton is assumed to be neglected and then
$g_{NL}$ is small. Now the cosmological observations become more and
more accurate, and even $g_{NL}$ can be detected in the near future
if it is not too small. So it is worth working out how $g_{NL}$ is
related to the curvaton self-interaction.

In \cite{Enqvist:2008gk,Enqvist:2005pg} the authors only considered
the case with small positive interaction term and they concluded
that $g_{NL}\sim {\cal O}(-10^4)-{\cal O}(-10^5)$ for the reasonable
cases. In this paper we will see that it is also reasonable to
consider the case with small negative interaction term for curvaton
model and discuss the dynamics of curvaton more carefully.

This paper is organized as follows. In Sec. 2, we discuss some
possible forms of the curvaton potential. In Sec. 3, the dynamics of
curvaton with potential slightly deviating from quadratic form is
studied in detail. In Sec. 4, we treat the non-quadratic potential
as a perturbation, and analytically calculate $f_{NL}$ and $g_{NL}$
to the leading order. How to determine the non-quadratic term from
experiments will be investigated in Sec. 5. At the end, we give a
summary in Sec. 6.

\setcounter{equation}{0}
\section{Estimator of small deviation from the quadratic potential}

We consider the following form of the curvaton potential,
\begin{equation}
  V(\sigma)=\frac{1}{2} m^2\sigma^2+\lambda
m^4 ({\sigma\over m})^n, \label{pt}
\end{equation}
The mass term is assumed to be dominant, and the size of the
interaction term compared with the mass term is measured by
\footnote{Here the value of $s$ is half of that in
Ref.\cite{Enqvist:2008gk,Enqvist:2005pg}.}
\begin{equation}
  s\equiv \lambda \left(\frac{\sigma_*}{m}\right)^{n-2},
\end{equation}
where $\sigma_*$ is the value of curvaton at Hubble exit. In
curvaton scenario the curvaton mass $m$ is assumed to be much
smaller than the Hubble parameter $H_*$ during inflation, and then
the value of curvaton at the end of inflation is also $\sigma_*$.

Usually the stability of the system requires the coupling constant
$\lambda$ to be positive. Otherwise the potential does not have a
lower bound when the vacuum expectation value of the field goes into
infinity. The authors in \cite{Enqvist:2008gk,Enqvist:2005pg} only
focused on the case of $s>0$. However we cannot expect the potential
is still reliable when $\sigma$ is very large. So we can expect that
$s$ can take a negative value.

On the other hand, to keep the mass of curvaton small enough, some
symmetries are called for to prevent the large quantum correction to
its mass. It seems natural to invoke supersymmetry. However
supersymmetry must be broken down during inflation and the mass
square of each scalar field generically receives a correction of
order $H_*^2$. An alternative candidate of curvaton is the
pseudo-Nambu-Goldstone boson \cite{Dimopoulos:2003az} --- axion. The
potential of axion is given by \e V(\sigma)=M^4\(1-\cos {\sigma\over
f}\), \q where $f$ is called the axion decay constant. The smallness
of the axion mass is protected by the shift symmetry
$\sigma\rightarrow \sigma+\delta$. In string compactifications,
axion fields are popular, and even when all other moduli are
stabilized, the axion potential remains rather flat as a consequence
of nonrenormalization theorems. In the inflation driven by axion,
the decay constant is required to be larger than the Planck scale.
However such a large value of $f$ cannot be achieved in string
theory \cite{ArkaniHamed:2003wu,Banks:2003sx}. But the axion is a
nice candidate of curvaton. For a small axion displacement
$\sigma<f$, the potential is expanded as \e V(\sigma)\simeq \half
{M^4\over f^2}\sigma^2-{1\over 24}{M^4\over f^4}\sigma^4+.... \q The
mass and the coupling of axion are respectively $m=M^2/f$ and
$\lambda=-M^4/(24f^4)$. In this case the parameter $s$ is \e
s=-{1\over 24}{\sigma_*^2\over f^2},\q which is negative. Since
$\sigma_*<f$, $-1/24<s<0$.

Because the curvaton mass is small compared to the Hubble parameter
$H_*$ during inflation, the Compton wavelength of curvaton is larger
than the background curvature radius $H_*^{-1}$, and the
gravitational effects may play a crucial role on the behavior of
curvaton. In \cite{Bunch:1978yq} the authors explicitly showed that
the quantum fluctuation of a light scalar field $\sigma$ with mass
$m$ in de Sitter space gives it a non-zero expectation value of
$\sigma^2$ as follows \e \langle\sigma^2\rangle={3H_*^4\over
8\pi^2m^2}. \q So the typical value of curvaton at the Hubble exit
is given by \e \sigma_*\simeq \sqrt{3\over 8\pi^2}{H_*^2\over m},\q
and then \e s=\lambda\(\sqrt{3\over 8\pi^2}{H_*^2\over
m^2}\)^{n-2}.\q For the axion-type curvaton, we have \e s=-{1\over
(8\pi)^2}{H_*^4\over m^2f^2}=-{1\over (8\pi)^2}{H_*^4\over M^4}. \q
Here we assume $\sigma_*<f$. Let's consider a concrete example of
axion in \cite{ArkaniHamed:2003wu} where a five-dimensional U(1)
gauge theory is compactified on a circle of radius $R$. A non-local
potential as a function of the gauge invariant Wilson loop \e
e^{i\theta}=e^{i\oint A_5 dx^5}\q will be generated in presence of
charged fields in the bulk. The effective Lagrangian for $\theta$
takes the form \e {\cal L}={1\over 2g_4^2(2\pi R)^2}(\p
\theta)^2-V(\theta), \q where the four-dimensional gauge coupling
$g_4$ is related to that in five dimensions by $g_4^2=g_5^2/(2\pi
R)$, and the potential is roughly given by \e V(\theta)\sim {c\over
16\pi^6}{1\over R^4}(1-\cos \theta), \q and $c\sim {\cal O}(1)$
depends on the number of charged fields in the bulk. In this case,
$f={1\over 2\pi g_4 R}$, $M^4={c\over 16\pi^6}{1\over R^4}$and the
parameter $s$ becomes \e s=-{\pi^4\over 4c}(H_*R)^4. \q In order for
the validity of the effective description of the four-dimensional
scenario, the size of extra dimension is required to be much smaller
than $H_*^{-1}$. If $s$ can be determined by experiments, we can
roughly know the ratio between the size of extra dimension and the
Hubble radius during inflation in this model.

\setcounter{equation}{0}
\section{The dynamics of curvaton after inflation}

After inflation radiation dominates our Universe and the Hubble
parameter drops with time as $H=1/(2t)$. The curvaton equation of
motion after inflation is \e \ddot \sigma+{3\over 2t}\dot
\sigma=-m^2\sigma\(1+n\lambda ({\sigma\over m})^{n-2}\). \q The
correction in the equation of motion is small if $|s|\ll 1/n$. We
introduce a dimensionless time coordinate $x=mt$ and the curvaton
equation of motion becomes \e \sigma''+{3\over
2x}\sigma'=-\sigma\(1+n\lambda ({\sigma\over m})^{n-2}\),
\label{s0e} \q here the prime denotes the derivative with respect to
$x$. The time when inflation ends is roughly given by
$t_{end}=1/(2H_*)$ and thus the initial time coordinate
$x_{ini}=mt_{end}\simeq 0$. Assume $x=x_o=mt_o$ at the time of
curvaton starting to oscillate. Perturbatively solving
Eq.(\ref{s0e}) with initial conditions $\sigma_{ini}=\sigma_*$ and
$\sigma_{ini}'=0$ at $x=0$, we find the value of curvaton at $x_o$
is \e \sigma_o=\sigma_*\[w(x_o)+ng(n,x_o)\lambda ({\sigma_*\over
m})^{n-2}+ {\cal O}(s^2)\], \label{ss}\q where \m
w(x_o)&=&2^{\frac{1}{4}}\Gamma\left(5/4\right)
  x_o^{-\frac{1}{4}} J_{\frac{1}{4}}(x_o), \\
g(n,x_o)&=&\pi
  2^{\frac{n-5}{4}}\Gamma(5/4)^{n-1}x_o^{-{1\over 4}} \nonumber \\
  &\times& \[J_{\frac{1}{4}}(x_o)\int_0^{x_o}
J_{\frac{1}{4}}^{n-1}(x)Y_{\frac{1}{4}}(x)x^{\frac{6-n}{4}}dx-
Y_{\frac{1}{4}}(x_o)\int_0^{x_o}
J_{\frac{1}{4}}^{n}(x)x^{\frac{6-n}{4}}dx \], \n and $J_{1/4}$ and
$Y_{1/4}$ are respectively the $1/4$ order Bessel $J$ function and
Bessel $Y$ function. The curvaton starts to oscillate when the
Hubble parameter drops below the same order of magnitude of the
curvaton mass $m$. Usually we have $H_o=m$ and then $x_o=\half$. In
\cite{Enqvist:2005pg} the authors assumed $x_o=1$. The values of
$w(x_o)$ and $g(n,x_o)$ for different $x_o$ are showed in Fig. 1.
\begin{figure}[h]
\begin{center}
\leavevmode \epsfxsize=0.45\columnwidth \epsfbox{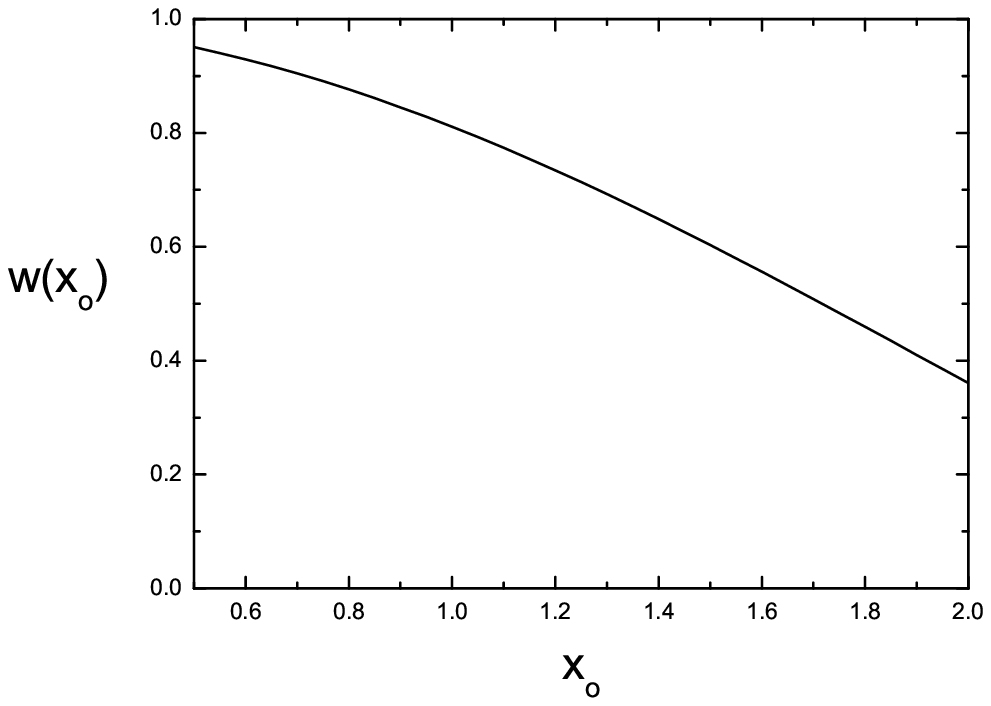} \leavevmode
\epsfxsize=0.45\columnwidth \epsfbox{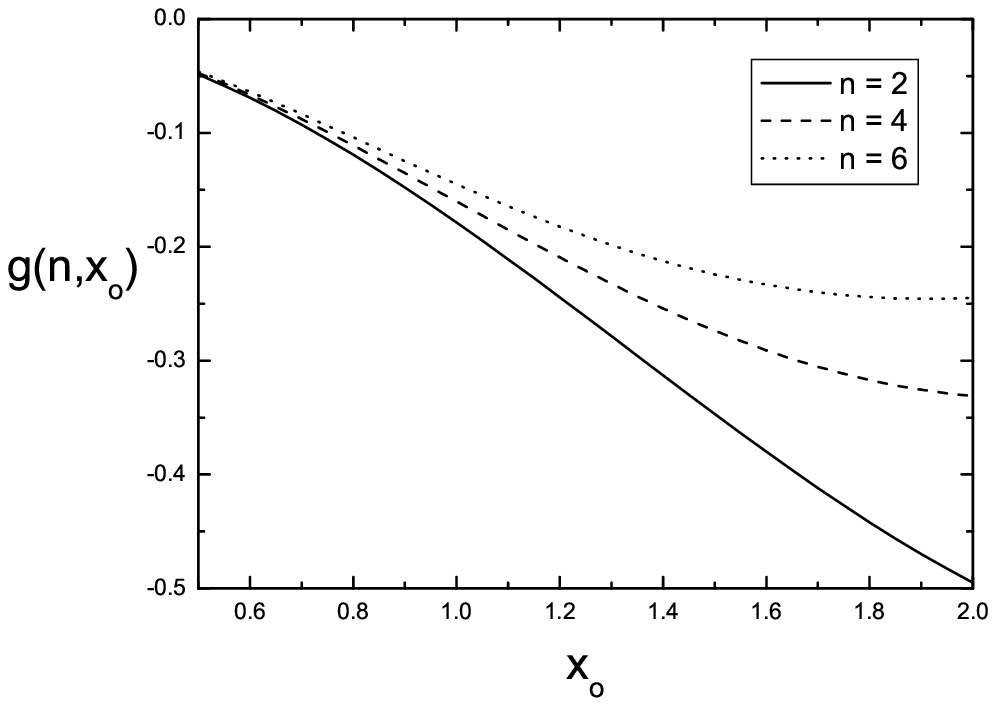}
\end{center}
\caption{The values of $w(x_o)$ and $g(n,x_o)$ for different $x_o$.}
\end{figure}
For $x_o=\half$, $w(x_o)\simeq 0.95$ and $g(2,x_o)\simeq -0.049$,
$g(4,x_o)\simeq -0.047$, $g(6,x_o)\simeq -0.046$. For $x_o=1$,
$w(x_o)\simeq 0.81$ and $g(2,x_o)\simeq -0.179$, $g(4,x_o)\simeq
-0.16$, $g(6,x_o)\simeq -0.145$. We see that $w(x_o)$ and $g(n,x_o)$
depend on the choice of the time when the curvaton starts to
oscillate.

\section{Non-linearity parameters in curvaton model}

The primordial curvature perturbation was calculated in
\cite{Linde:1996gt,Lyth:2002my}. The primordial power spectrum,
bispectrum and trispectrum are defined by \m \langle\zeta({\bf
k_1})\zeta({\bf k_2})\rangle&=&(2\pi)^3 {\cal
P}_{\zeta}(k_1)\delta^3({\bf k_1}+{\bf k_2}), \\ \langle\zeta({\bf
k_1})\zeta({\bf k_2})\zeta({\bf k_3})\rangle&=&(2\pi)^3 B_\zeta(k_1,k_2,k_3)\delta^3({\bf k_1}+{\bf k_2}+{\bf k_3}), \\
\langle\zeta({\bf k_1})\zeta({\bf k_2})\zeta({\bf k_3})\zeta({\bf
k_4})\rangle&=&(2\pi)^3 T_\zeta(k_1,k_2,k_3,k_4)\delta^3({\bf
k_1}+{\bf k_2}+{\bf k_3}+{\bf k_4}). \n The bispectrum and
trispectrum are respectively related to the power spectrum by \m
B_\zeta(k_1,k_2,k_3)&=&{6\over 5}
f_{NL}[{\cal P}_\zeta(k_1){\cal P}_\zeta(k_2)+2\ \hbox{perms}], \\
T_\zeta(k_1,k_2,k_3,k_4)&=&\tau_{NL}[{\cal P}_\zeta(k_{13}){\cal
P}_\zeta(k_3){\cal P}_\zeta(k_4)+11\ \hbox{perms}] \nonumber \\
&+&{54\over 25}g_{NL}[{\cal P}_\zeta(k_2){\cal P}_\zeta(k_3){\cal
P}_\zeta(k_4)+3\ \hbox{perms}]. \n Here the non-linearity parameter
$\tau_{NL}$ depends on $f_{NL}$: \e \tau_{NL}={36\over 25}f_{NL}^2.
\q In curvaton model, the amplitude of the primordial power spectrum
and the non-linearity parameters can be written as \m
P_\zeta&=&{1\over 9\pi^2}f_D^2q^2{H_*^2\over \sigma_*^2}, \\
f_{NL}&=&{5\over 4f_D}\(1+h\)-{5\over 3}-{5f_D\over 6}, \label{fnl} \\
g_{NL}&=& {25\over 54}\[{9\over 4f_D^2}({\tilde h}+3h)-{9\over
f_D}(1+h)+\half (1-9h)+10f_D+3f_D^2 \right], \label{gnl} \n where \m
f_D&=&{3\Omega_{\sigma,D}\over 4-\Omega_{\sigma,D}}, \quad
q={\sigma_*\sigma_o'\over \sigma_o}, \\ h&=&{\sigma_o\sigma_o''\over
{\sigma_o'}^2}, \quad {\tilde h}={\sigma_o^2\sigma_o'''\over
{\sigma_o'}^3}. \n According to Eq.\eqref{ss}, the parameters $q$,
$h$ and $\tilde h$ take the form
\m q&=&{w(x_o)+n(n-1)g(n,x_o)s\over w(x_o)+ng(n,x_o)s}, \\ \label{hexplicit}
h&=&{w(x_o)+ng(n,x_o)s\over (w(x_o)+n(n-1)g(n,x_o)s)^2}n(n-1)(n-2)g(n,x_o)s,\\
{\tilde h}&=&{(w(x_o)+ng(n,x_o)s)^2 \over
(w(x_o)+n(n-1)g(n,x_o)s)^3}n(n-1)(n-2)(n-3)g(n,x_o)s. \label{h}
\label{tildehexplicit}\n We see that both $h$ and $\tilde h$ are
proportional to the size of the non-quadractic term $s$. If the
curvaton potential deviates from the exactly quadractic form, the
value of $f_{NL}$ can be small even when $f_D\ll 1$ if 1 is
cancelled by $h$, but $g_{NL}$ is still very large.

Here we also want to stress that the choice of the time when
curvaton starts to oscillate has much effect on the quantitative
results. For $n=4$, we illustrate the quantitative results of $q$,
$h$ and $\tilde h$ for different choice of $x_o$ in Fig. 2.
\begin{figure}[h]
\begin{center}
\leavevmode \epsfxsize=0.45\columnwidth \epsfbox{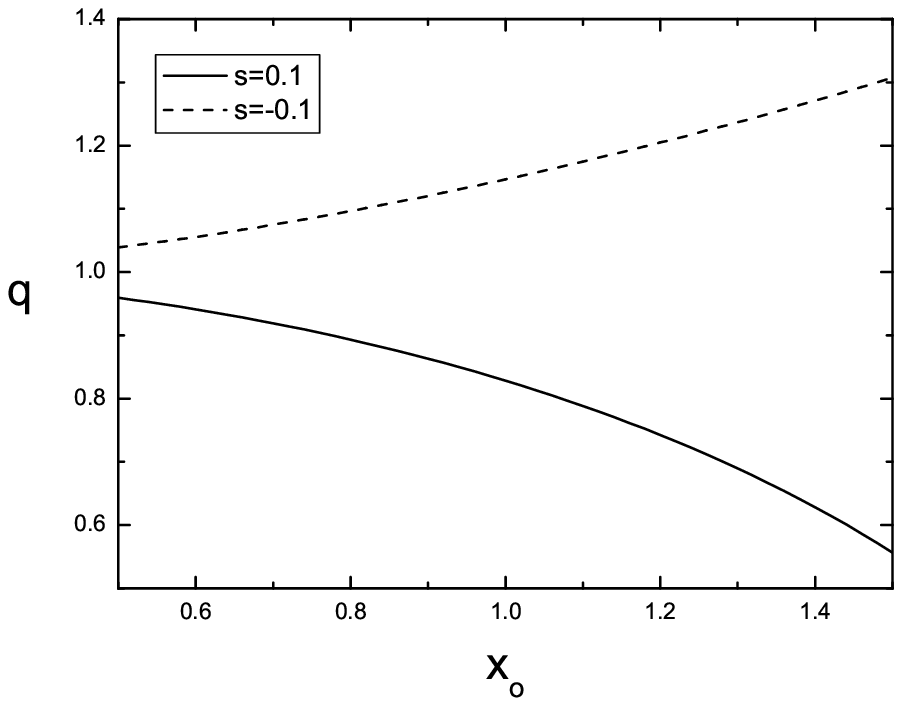} \\
\leavevmode \epsfxsize=0.45\columnwidth \epsfbox{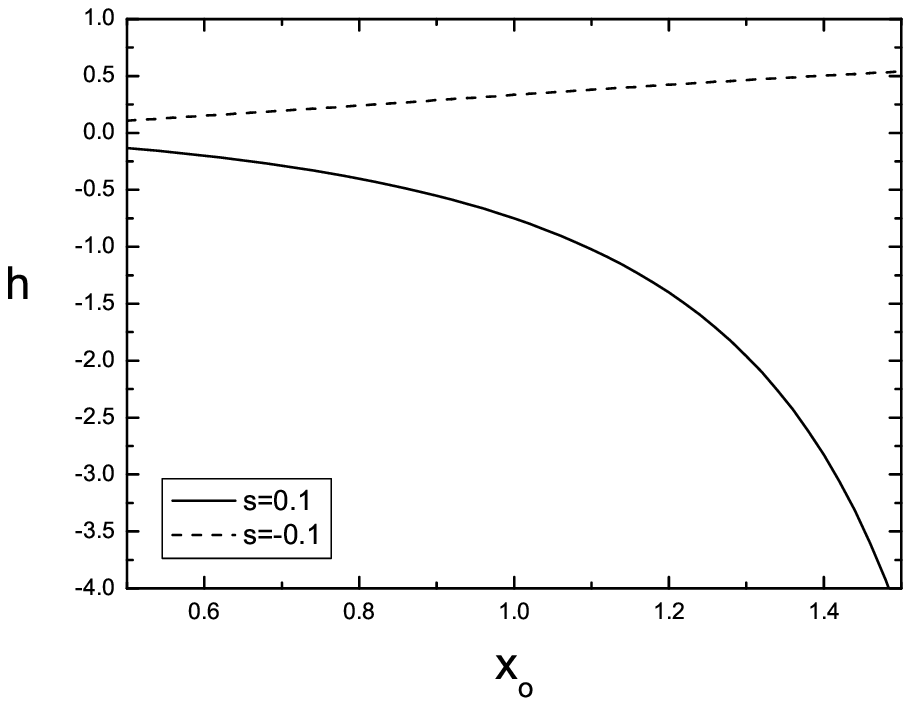} \leavevmode
\epsfxsize=0.45\columnwidth \epsfbox{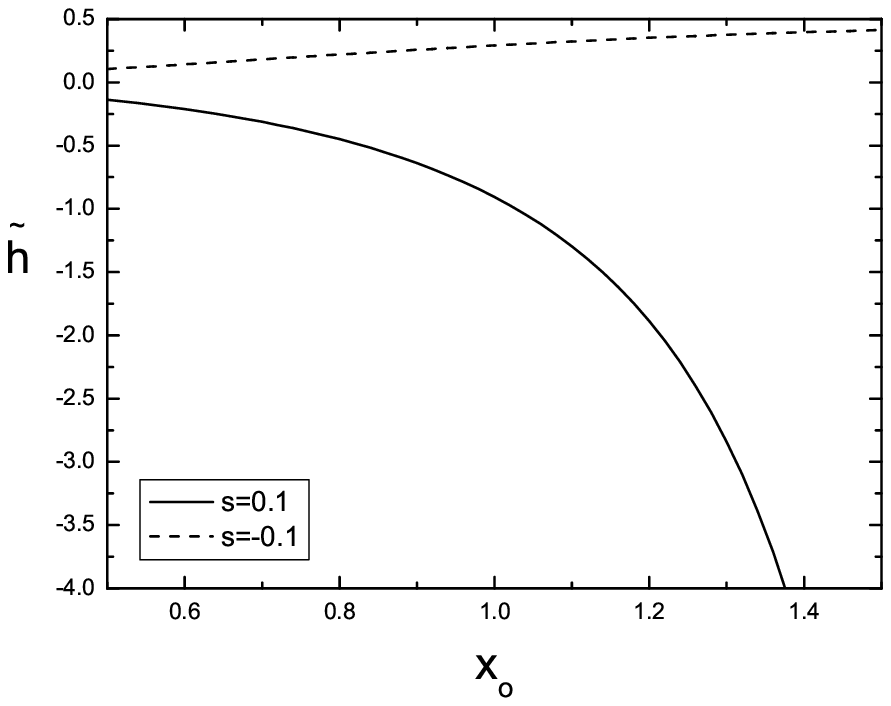}
\end{center}
\caption{The quantitative results of $q$, $h$ and $\tilde h$ for
different $x_o$. Here we take $n=4$. }
\end{figure}

In \cite{Enqvist:2008gk} the authors only consider the case of $s>0$
and they concluded that $f_{NL}$ can be small, but $g_{NL}$ will be
${\cal O}(-10^4)-{\cal O}(-10^5)$ for the reasonable cases. In Sec.
2 we argued that it is reasonable to extend the discussions into the
case of $s<0$. For $s<0$, both $h$ and $\tilde h$ are positive,
$f_{NL}$ always gets a large positive value if $f_D\ll 1$, and
$g_{NL}\sim f_{NL}^2$ which is roughly the same order of magnitude
as $\tau_{NL}$. See the numerical results of $f_{NL}$ and $g_{NL}$
in Fig. 3.
\begin{figure}[h]
\begin{center}
\leavevmode \epsfxsize=0.45\columnwidth \epsfbox{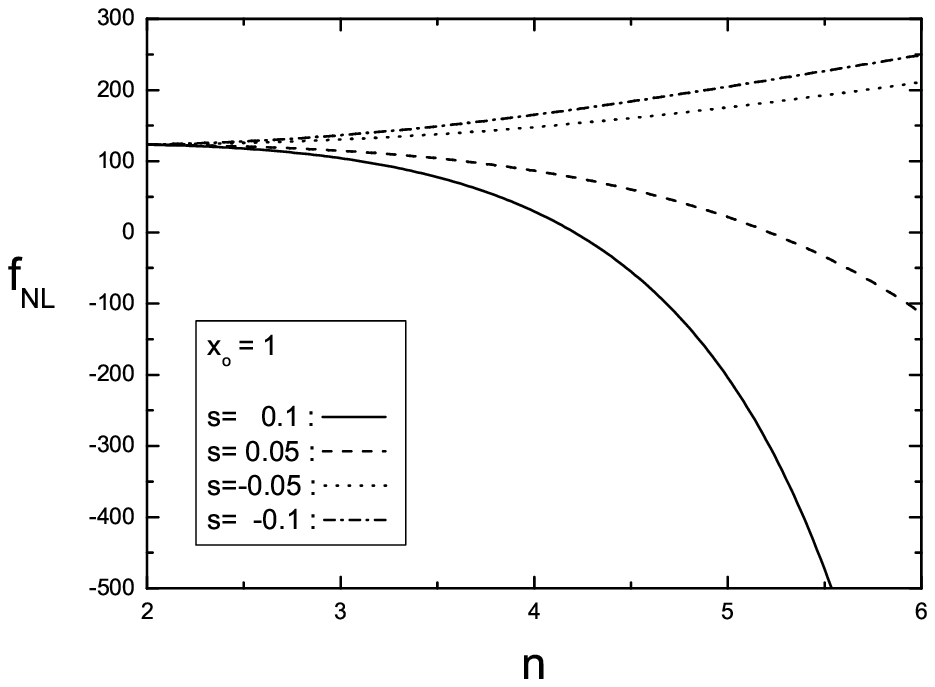}
\leavevmode \epsfxsize=0.45\columnwidth \epsfbox{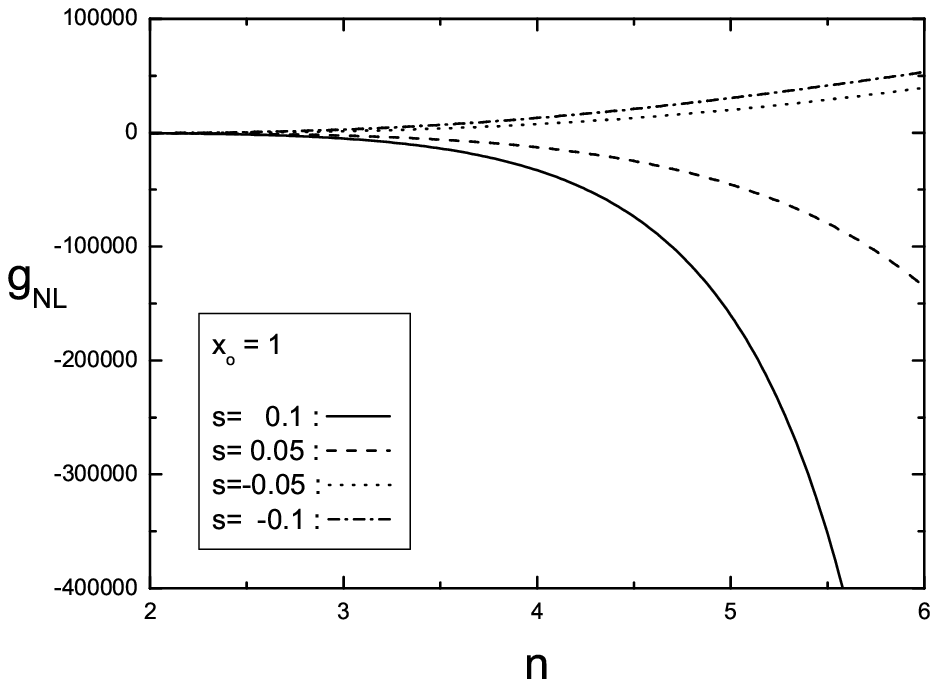}
\leavevmode \epsfxsize=0.45\columnwidth \epsfbox{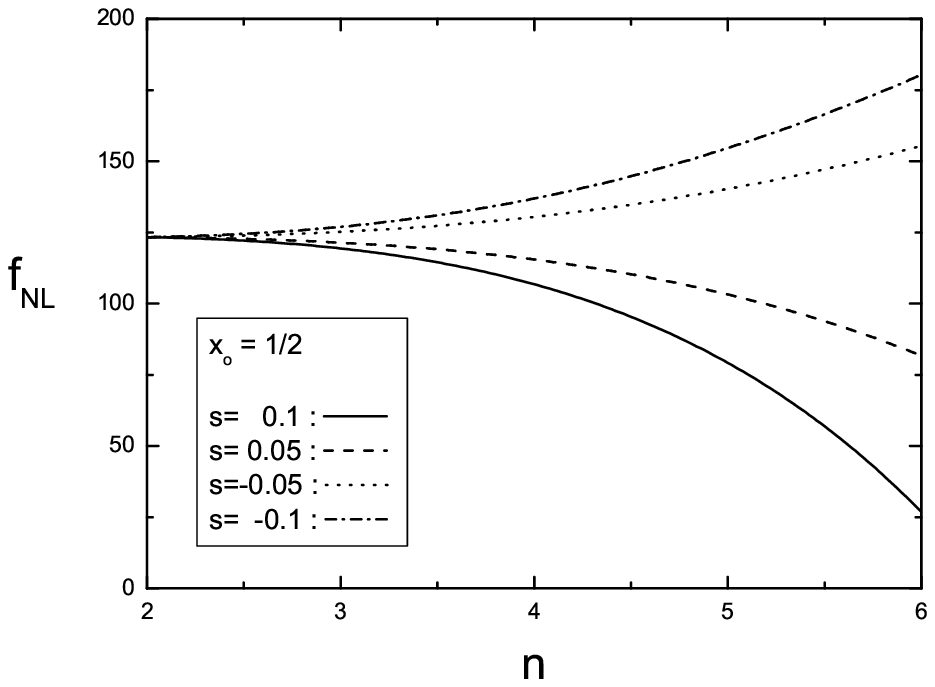}
\leavevmode \epsfxsize=0.45\columnwidth \epsfbox{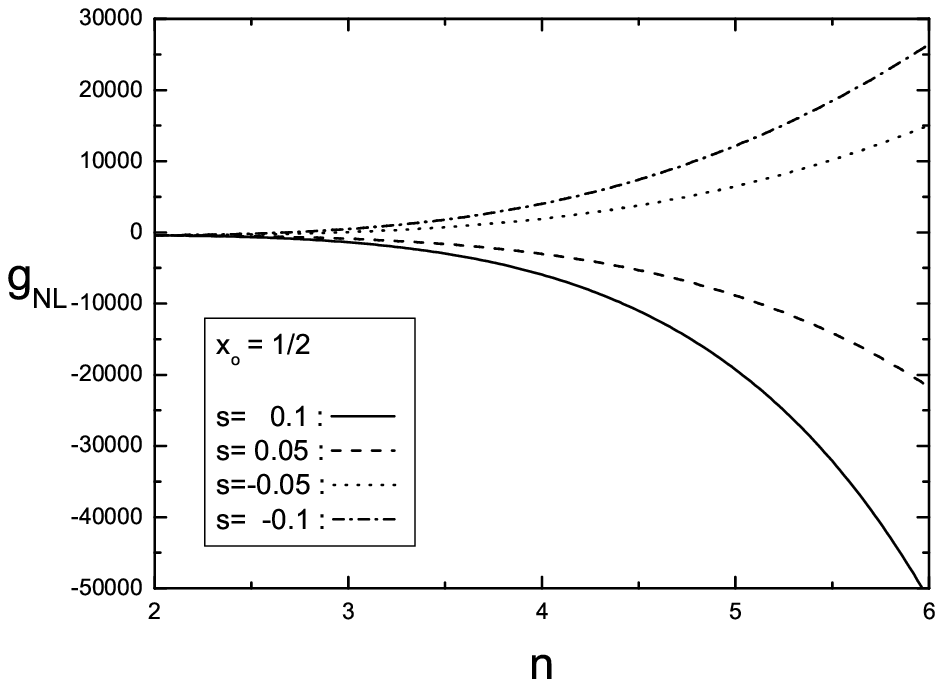}
\end{center}
\caption{The non-linearity parameters $f_{NL}$ and $g_{NL}$ for
different $n$, $s$ and $x_o$ are plotted. Here we take $f_D=0.01$. }
\end{figure}

\setcounter{equation}{0}
\section{A proposal to measure the non-quadratic term}

In \cite{Enqvist:2008gk} the authors pointed out that any deviation
from the relation \e g_{NL}\simeq -{10\over 3}f_{NL}\q will indicate
that the curvaton potential does not take the exactly quadratic
form. Here we will take a closer look at the measurement of the
curvaton self-interaction term.

From Eq.\eqref{fnl} and \eqref{gnl}, we find
\begin{equation}
  g_{NL}+\frac{10}{3}f_{NL}=\frac{25}{24} \frac{\tilde
  h+3h}{f_D^2}+{25\over 108}(-23-9h+8f_D+6f_D^2).
\end{equation}
Here we only focus on the case with large absolute values of
$f_{NL}$ and $g_{NL}$, because the accuracy of experiments is
limited. Or equivalently, we have $f_D\ll 1$. If $|s|\ll 1$, we have
$f_{NL}\simeq{5\over 4f_D}$ and \e g_{NL}+{10\over 3}f_{NL}\simeq
{2\over 3w(x_o)}f_{NL}^2n^2(n-1)(n-2)g(n,x_o)s. \q For convenience,
we define a new quantity $\alpha(s)$ as follows \e \alpha(s)\equiv
{w(x_o)(3g_{NL}+10f_{Nl})\over 2f_{NL}^2n^2(n-1)(n-2)g(n,x_o)}.
\label{als}\q If $|s|\ll 1$, we can expect $\alpha(s)=s$. Since
there are three free parameters ($f_D$, $n$, $s$) from the
theoretical viewpoint, but only two observed quantities ($f_{NL}$,
$g_{NL}$), we cannot model-independently determine the value of $s$
from experiments. For an instance, we consider the interaction term
takes the form $\lambda \sigma^4$ and show $\alpha(s)$ as a function
of $s$ in Fig. 4.
\begin{figure}[h]
\begin{center}
\leavevmode \epsfxsize=0.8\columnwidth \epsfbox{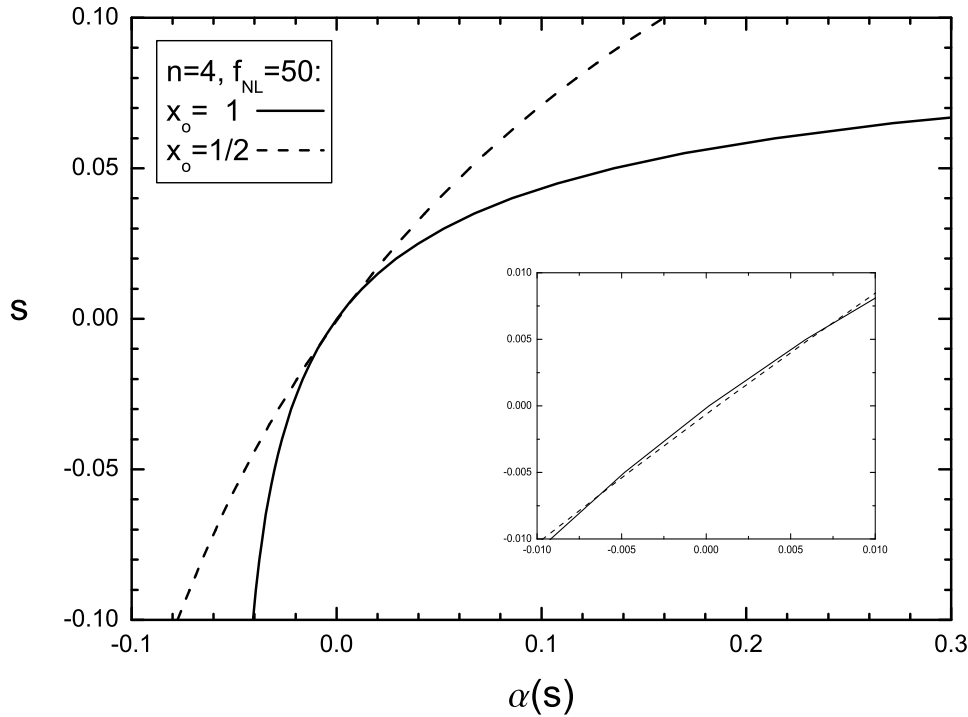}
\end{center}
\caption{This figure illustrates $\alpha(s)$ as a function of $s$.
In fact, for the large value of $f_{NL}$, it is insensitive to
$f_{NL}$.}
\end{figure}
Even though we set $f_{NL}=50$, one can check that $\alpha(s)$ is
quite insensitive to $f_{NL}$. In this case $\alpha(s)\simeq s$ if
$|s|<0.01$ and this result is roughly independent on the choice of
$x_o$.

Before the end of this section, we also want to estimate the order
of magnitude of the dimensionless coupling coupling $\lambda$ in the
case of $n=4$. The typical value of curvaton during inflation is
$\sigma_*\sim H_*^2/m$, and then $s\sim \lambda (H_*/m)^4$ and the
amplitude of the primordial power spectrum is $P_\zeta\sim
f_D^2m^2/H_*^2$. For large $f_{NL}$ and small $s$, we have
$f_{NL}\sim 1/f_D$. Therefore $m/H_*\sim \sqrt{P_\zeta}f_{NL}$ and
$\lambda \sim s(\sqrt{P_\zeta}f_{NL})^4$. Taking into account WMAP
normalization $P_\zeta=2.457\times 10^{-9}$ \cite{Komatsu:2008hk},
$\lambda\sim 10^{-18}f_{NL}^4s$. For $f_{NL}\sim 10^2$ and $s\sim
\pm 10^{-2}$, $g_{NL}\sim -10sf_{NL}^2\sim \mp 10^3$ and
$\lambda\sim \pm 10^{-12}$. Such a weak interaction can be possibly
detected by the cosmological experiments in the near future!

\setcounter{equation}{0}
\section{Conclusions}

The fluctuations of curvaton field evolve non-linearly on the
superhorizon scales if the curvaton potential deviates from the
exactly quadratic form. In this paper, we suggest that the leading
order of the non-quadratic term for curvaton field can be negative,
for example in the axion-type curvaton model. We also investigate
the curvaton dynamics and the non-linearity parameters in curvaton
model with non-quadratic curvaton potential in detail. For a
positive non-quadratic term, $f_{NL}$ can be very small even when
$f_D\ll 1$. But it does not happen in the case with a negative
non-quadratic term. We also see that the second order non-linearity
parameter $g_{NL}$ measures the size of the self-interaction of
curvaton field. If $g_{NL}$ is positive, the leading order of the
self-interaction term should be negative, and the axion-type
curvaton model is preferred. The next generation of experiments such
as Planck will improve the accuracy to $\Delta f_{NL}\sim 6$ and
$\Delta\tau_{NL}\sim 560$ \cite{Kogo:2006kh}. We hope the
non-linearity parameters will be detected soon.

In this paper, the self-interaction term is treated as a
perturbation. The case in which the self-interaction term dominates
the curvaton potential during inflation will be discussed in
\cite{Huang:2008zj}. On the other hand, multiplicity of curvaton
fields is expected in the fundamental theories going beyond the
standard model. In particular, axions are typically present in large
numbers in string compactifications. In \cite{Huang:2008rj}, these
axion fields are suggested to be taken as curvatons. It is worth
investigating the non-linearity parameters in N-vaton model
\cite{Huang:2008rj} carefully in the future.

\vspace{1cm}

\noindent {\bf Acknowledgments}

We would like to thank E.~J.~Chun and M. Li for useful discussions.

\end{document}